\documentclass[journal]{IEEEtran}
\pdfoutput=1

\renewcommand{\baselinestretch}{1.0}

\usepackage{verbatim} 
\usepackage[utf8]{inputenc}
\usepackage[hidelinks,breaklinks]{hyperref}
\usepackage{multirow}
\usepackage{dingbat}
\usepackage[figuresleft]{rotating}
\usepackage{tablefootnote}

\title{A Survey of Published Attacks on Intel SGX}
\IEEEoverridecommandlockouts%
\author{
	\IEEEauthorblockN{
		\thanks{
			This work was partially supported by the Wallenberg AI, Autonomous Systems and Software Program (WASP) funded by the Knut and Alice Wallenberg Foundation
		}
		Alexander Nilsson\IEEEauthorrefmark{1}\IEEEauthorrefmark{2}, 
		Pegah Nikbakht Bideh\IEEEauthorrefmark{1}, 
		Joakim Brorsson\IEEEauthorrefmark{1}\IEEEauthorrefmark{3}\IEEEauthorrefmark{4}
	}
	\IEEEauthorblockA{\texttt{\{alexander.nilsson,pegah.nikbakht\_bideh,joakim.brorsson\}@eit.lth.se}}\\
	\IEEEauthorblockA{\IEEEauthorrefmark{1}Lund University, Department of Electrical and Information Technology, Sweden}\\
	\IEEEauthorblockA{\IEEEauthorrefmark{2}Advenica AB, Sweden}\\
	\IEEEauthorblockA{\IEEEauthorrefmark{3}Combitech AB, Sweden}\\
	\IEEEauthorblockA{\IEEEauthorrefmark{4}Hyker Security AB, Sweden}
}
\newcommand{\secref}[1]{sec.~\ref{#1}}

\usepackage[backend=bibtex,style=ieee]{biblatex}
\usepackage{graphicx}
\usepackage{bytefield}
\usepackage{booktabs}

\usepackage[nointegrals]{wasysym}
\usepackage{array}
\newcolumntype{H}{>{\setbox0=\hbox\bgroup}c<{\egroup}@{}}

\usepackage[obeyDraft,colorinlistoftodos,textwidth=2cm]{todonotes}
\usepackage[fit]{truncate}
\usepackage{setspace}
\usepackage{xparse}
\usepackage{soul}
\usepackage{amsmath}

\makeatletter
\if@todonotes@disabled%

\else

\fi
\makeatother

\reversemarginpar%
\definecolor{bleudefrance}{rgb}{0.19, 0.55, 0.91}
\newcommand{\TODOIMPL}[3]{\todo[#3,caption={\truncate{15cm}{\textbf{#2}: #1}}]{\renewcommand{\baselinestretch}{0.75}\selectfont\textbf{#2}: #1\par}}
\DeclareDocumentCommand{\TODO}{O{Add text here}}{\TODOIMPL{#1}{Todo}{inline,backgroundcolor=orange}}
\DeclareDocumentCommand{\FIX}{O{needed}}{\TODOIMPL{#1}{Fix}{inline,backgroundcolor=red,size=\footnotesize}}
\DeclareDocumentCommand{\MISSREF}{O{missing}}{\TODOIMPL{#1}{Reference}{backgroundcolor=white,fancyline,size=\footnotesize}}
\DeclareDocumentCommand{\INVESTIGATE}{O{this}}{\TODOIMPL{#1}{Investigate}{inline,backgroundcolor=bleudefrance}}
\DeclareDocumentCommand{\ADDCOMMENT}{O{here}}{\TODOIMPL{#1}{Comment}{inline,backgroundcolor=yellow}}
\DeclareDocumentCommand{\REVIEWTHIS}{O{please}}{\TODOIMPL{#1}{Review this}{inline,backgroundcolor=green}}
\DeclareDocumentCommand{\REVIEWER}{O{here}}{\TODOIMPL{``#1''}{Reviewer comment}{inline,backgroundcolor=red}}
\DeclareDocumentCommand{\REVIEWERDONE}{O{here}}{\TODOIMPL{\st{``#1''}}{Fixed! \st{Reviewer comment}}{inline,backgroundcolor=white}}

\newcommand{\citeNA}[1]{\IEEEnoauxwrite{\cite{#1}}}

\newcommand{\MatrixYes}{\CIRCLE}
\newcommand{\MatrixYesC}[1]{\CIRCLE\textsuperscript{\citeNA{#1}}}
\newcommand{\MatrixYesCC}[2]{\CIRCLE\(^{\text{\citeNA{#1}}}_{\text{\citeNA{#2}}}\)}

\newcommand{\MatrixPartly}{\LEFTcircle}
\newcommand{\MatrixPartlyC}[1]{\LEFTcircle\textsuperscript{\citeNA{#1}}}

\newcommand{\MatrixNo}{\Circle}

\newcommand*\rot{\rotatebox{90}}

\newcommand{\AttackControlledChannel}{\secref{sec:controlled_channel}}
\newcommand{\AttackCache}{\secref{sec:cache_attacks}}
\newcommand{\AttackSpeculative}{\secref{sec:attacks_spec_exec}}
\newcommand{\AttackBranchPredict}{\secref{sec:branch_prediction}}
\newcommand{\AttackRogueDataCacheLoad}{\secref{sec:rogue_data_cache_loads}}
\newcommand{\AttackMDS}{\secref{sec:attacks_mds}}
\newcommand{\AttackSoftFaultInject}{\secref{sec:soft-fault-injection}}

\newcommand{\DefenceTypeMicroCode}{Microcode patch}
\newcommand{\DefenceTypeCompiler}{Compiler/SDK}
\newcommand{\DefenceTypeApplication}{Application design} 

\newcommand{\DefenceTypeSystem}{System design}

\newcommand{\kw}[1]{\textsf{#1}}

\IEEEaftertitletext{%
}

\begin{document}

\maketitle
\begin{abstract}
	Intel Software Guard Extensions (\kw{SGX}) provides a trusted execution environment (\kw{TEE}) to run code and operate sensitive data.
	\kw{SGX} provides runtime hardware protection where both code and data are protected even if other code components are malicious.
	However, recently many attacks targeting \kw{SGX} have been identified and introduced that can thwart the hardware defence provided by \kw{SGX}.
	In this paper we present a survey of all attacks specifically targeting Intel \kw{SGX} that are known to the authors, to date.
	We categorized the attacks based on their implementation details into 7 different categories.
	We also look into the available defence mechanisms against identified attacks and categorize the available types of mitigations for each presented attack.
\end{abstract}

\section{Introduction}\label{sec:introduction}

\TODO[write the history of TEE and how it led up to SGX]
\TODO[introduce side-channel attacks and their history]
\TODO[write reasoning for categorization]
\TODO[put a contribution (first survey, new categorization, etc)]
\TODO[compare impact of side-channel attacks on sgx with TPM and VMs?]
Trusted Execution Environments (\kw{TEEs}) create isolated environments where sensitive code can run with higher security level than the operating system.
Intel Software Guard Extensions (\kw{SGX}) is an example of a \kw{TEE}.
\kw{SGX} utilizes enclaves to isolate execution environment from other applications, the operating system's kernel and the hypervisor.
\kw{SGX} can run arbitrary code on general hardware and is suitable for cloud environments where it isolates the running code and data from the untrusted environment.

\TODO[start this paragraph with description of need of isolation and attestation, then say VM and TPM, then say SGX better]
Without \kw{TEE} solutions such as \kw{SGX}, virtualization techniques are the primary defence that can be used by software to isolate code and data from other running software on a computer.
Unfortunately virtualization techniques requires the application to trust the \kw{OS}-kernel and hypervisor, and by extension the cloud provider in such a scenario.

\kw{SGX} is only one of a few attempts at solving the issue of trusted computing in the cloud. Another solution is the Trusted Platform Module (\kw{TPM}).
The \kw{TPM} however requires a larger chain of trust which is a drawback since it would require that the user roots its trust in the both the intentions of the implementers and in the absence of bugs in very large pieces of code (often including the BIOS', the OS kernel's and hypervisor's code-bases).
Comparatively, \kw{SGX} provides a great advantage, in that the root of trust is based only on the application code itself and the hardware implementation of the CPU\@.

Unfortunately, a relatively large number of flaws and attacks against \kw{SGX} have been published by researchers over the last few years.

\subsection{Contribution}
In this paper, we present the first comprehensive review that includes all known attacks specific to \kw{SGX}, including controlled channel attacks, cache-attacks, speculative execution attacks, branch prediction attacks, rogue data cache loads, microarchitectural data sampling and software-based fault injection attacks.
For most of the presented attacks, there are countermeasures and mitigations that have been deployed as microcode patches by Intel or that can be employed by the application developer herself to make the attack more difficult (or impossible) to exploit.
For all of the surveyed attacks in this paper, any known and relevant mitigation techniques are also presented.

\subsection{Organization}
In \secref{sec:background} some background information on \kw{SGX} is presented.
The known attacks with their categorizations are given in \secref{sec:sgx-attacks}. Then, the available mitigation techniques to categorized attacks are given in \secref{sec:mitigate}.
Finally, the current status of mitigation techniques and their applicability against specific attacks are discussed in \secref{sec:status} and the paper is concluded in \secref{sec:conclusions}.

\section{Background on \kw{SGX}}\label{sec:background}

\kw{SGX} is a set of extensions that aim to provide integrity and confidentiality for secure computations on computer systems where privileged software is potentially malicious.

\kw{SGX} provides execution environments called enclaves to run code and operate sensitive data, where both code and data are protected from the outside software environment.
This includes other applications running on the system and the operating system's kernel.
Even the hypervisor, if it is running, is an actor from which \kw{SGX} enclaves are protected.
Notably, physical attacks are not considered in Intel's threat model, nor are so-called side-channel attacks.

For the rest of this section we refer to~\cite{costan2016intel} without explicitly writing it out on each paragraph. We refer to it also for the interested reader who wishes a more detailed explanation on the internals of \kw{SGX}.

\subsection{\kw{SGX} Overview}

The Intel x86 64-bit instruction set architecture (\kw{ISA}), to which we limit the scope of this paper, defines several architectural privilege levels, each one strictly more capable than the one below it.
The least privileged is ring 3 where all user-space applications run.
This is the majority of the software on a running system\footnote{This includes large parts of the \kw{OS} as well, where the kernel is the obvious exception.}.
Ignoring ring 2 and 1, which are not used by any major Operating System today, the next-most powerful privilege level is ring 0 in which the \kw{OS} kernel is running.
Software running in ring 0 is responsible for resource allocation, device management, context switching, page swapping and so forth.

Intel Virtual Machine Extensions (\kw{VMX}) is a set of hardware virtualization instructions which introduces the additional privilege levels of \kw{VMX} root and \kw{VMX} non-root.
Hypervisors usually run as \kw{VMX} root in ring 0 and carry the ultimate responsibility of resource allocation.


This relates to Intel \kw{SGX} in the following way: An \kw{SGX} enclave always runs as ring 3 like any normal user-space application (either \kw{VMX} root or non-root).
Also like any normal user-space application it relies on the \kw{OS}-kernel (ring 0) software for services such as scheduling, page swapping and hardware interrupt handling.
This is despite the fact that none of the system software (ring 3 or 0) is trusted by the enclave threat model.
This has been achieved by a rather complex series of hardware extensions as well as the exclusion of denial-of-service from the threat model.
This is reasonable since protection against denial of service in an untrusted environment would be very hard to achieve, if not down-right impossible.

The code and data for all enclaves on a running system resides in the \emph{Enclave Page Cache} (\kw{EPC}) inside the \emph{Processor Reserved Memory} (\kw{PRM}) which is a reserved subset of the physical address space (\kw{DRAM}).
It is worth noting that this address range is protected by the \kw{CPU} so that Direct Memory Access (\kw{DMA}) is prohibited and that not even code running in the so-called Software Management Mode\footnote{
	An even higher privilege level than ring 0 and \kw{VMX}-root, used solely by the motherboard firmware to manage, for example, the booting stage, fan control, power and sleep functions.
} can get access to its contents.
In order to protect against snoops of external memory reads and writes the \kw{PRM} is transparently encrypted and integrity protected before entering/exiting the memory bus.
This means that the \kw{CPU} package itself is the only place where enclave data can be read in its decrypted form.

Enclaves are designed to operate much like dynamic loadable modules%
\footnote{Such as \kw{.dll} files for \kw{PE} and Windows based systems and \kw{.so} files for \kw{ELF} and Unix derived systems.}
which are loaded directly into the virtual address space of user-space applications.
This means that enclaves can be entered in much the same ways that \kw{API}-calls are made into software libraries (although it is a more expensive operation).
This makes it comparatively easy to modify existing programs.

Enclaves can only be entered at well defined entry points (much like a library \kw{API}) as specified by the enclave author.
This prevents memory mapping attacks and security check bypasses.
While application software cannot access the memory space of the enclave the reverse is not true.
The enclave have no restrictions in regards to the rest of the applications code and data, this facilitates easy and secure communication between the two modes.

The \kw{SGX} design and implementation is fully backward compatible with other \kw{ISA} extensions such as \kw{VMX} which enables the use of this technology by cloud tenants where several virtual machines are co-hosted on the same hardware.

\subsection{The \kw{SGX} Lifetime}

The enclave's lifetime is managed by the (untrusted) \kw{OS}-kernel, this includes handling of page swapping, interrupts and \kw{CPU} core scheduling.
This is facilitated by several new instructions introduced by the \kw{SGX} extensions.
Some of the more important ones will be discussed in this section.

\subsubsection{Creation}
In order to create an enclave, ring 0 first issues the privileged \kw{ECREATE} instruction.
Enclave creation is intended as a service for applications, provided by the system software.
The \kw{ECREATE} instruction allocates a special page for the enclave called the \kw{SECS}, like all other enclave pages it is located inside the protected \kw{PRM} range.
The \kw{SECS} stores meta data for the enclave and it is critical for the enclave's security.

\subsubsection{Loading}
After creation the \kw{SECS} is still marked as \emph{uninitialized}.
Only while the \kw{SECS} is marked as such can \kw{EADD} and \kw{EEXTEND} instructions be issued for that enclave.
These instructions are also privileged and can only be issued by ring 0.
\kw{EADD} is used to add pages into the protected virtual address space of the enclave.

\kw{EEXTEND} is used for measuring data and code for software attestation.
Attestation will be briefly discussed in \secref{sec:attestation}.

\subsubsection{Initialization}
The \kw{OS}-kernel in ring 0 must issue the \kw{EINIT} instruction in order to initialize the enclave.
However, before it can do that it must first obtain an \kw{EINIT} Token Structure.
The procedure for this is to utilize a special \emph{Launch Enclave} (\kw{LE}) which is signed by a special key whose corresponding public part is hardcoded into the \kw{SGX} implementation by Intel.

\subsubsection{Teardown}
Ring 0 can issue the \kw{EREMOVE} instruction to remove enclaves.
This deallocates the specified page after it is made sure that no logical processor currently owns it.
After the \kw{SECS} page is deallocated the enclave is completely destroyed.
\kw{EREMOVE} refuses to deallocate the \kw{SECS} before all other pages have been deallocated.

\subsubsection{Synchronous Entry}\label{sec:syncenter}
Each logical processor executing the enclave code uses a Thread Control Structure (\kw{TCS}) which controls the execution and makes sure that no two processors use the same \kw{TCS} at the same time.

Disregarding the possibility of interrupts an enclave is executed as a controlled jump into the enclave's code by issuing the \kw{EENTER} instruction.
\kw{EENTER} can only jump to predefined addresses which prevents a malicious host application from bypassing security checks that the enclave author might wish to perform.
When entering enclave mode some registers are saved in order to be restored later when the enclave is done executing. While executing enclave code, the logical processor is said to be in \emph{enclave mode}.

\subsubsection{Synchronous Exit}
One of two ways of exiting enclave mode is via the \kw{EEXIT} instruction which additionally performs a restore of the registers saved by \kw{EENTER}.

\subsubsection{Asynchronous Exit}
If a hardware exception occurs, such as an interrupt or fault while a logical processor is executing enclave code an \kw{AEX} instruction is issued by the enclave before invoking the system software's default exception handler.
This instruction saves the current execution context and restores the state saved by \kw{EENTER}.


\subsubsection{Resumption}\label{sec:asyncenter}
Once the registered software handler for a hardware interrupt has finished, it jumps back to the asynchronous exit handler in the enclaves host's process.
This handler is responsible for issuing the \kw{ERESUME} instruction which puts the logical processor back into enclave mode which continues the execution which was interrupted.

\subsubsection{Page Eviction and Reloading}


The \kw{SGX} eviction implementation relies on the privileged \kw{EWB} instruction (restricted to ring 0) which encrypts and integrity protects the specified page with a symmetric key known only to the enclave.
It also utilizes a mechanism for ensuring freshness, based on nonces. After the \kw{EWB} instruction has evicted the \kw{PRM} page, the system's default page-swapping mechanism can take over and flush the page to disk, if desired.

\subsection{Software Attestation}\label{sec:attestation} The goal of software attestation is to verify that the software application running inside an enclave is trustworthy.
This verification can be done by using a remote party.
Attestation data within the software can be signed by requesting the \kw{SGX} hardware implementation to generate an attestation signature.
The signature can be used to uniquely identify the software and any optional data inside the enclave.
The verifier can use the signature to make sure that the attestation data was generated by a specific software running on a genuine \kw{SGX} implementation.

\section{Attacks on SGX}\label{sec:sgx-attacks}

We found a large group of attacks on \kw{SGX} (most of them side-channel based) and using their implementation details, we divide them into 7 different categories.
The categories and the attacks in each category are described below. The surveyed papers are mostly self-categorized, we have simply unified the most common terminology based on their technical details.
A summary of the attacks presented in this section is provided in Table~\ref{tab:attack_matrix}.

\paragraph{Attacks that are out-of-scope}
As mentioned, we consider denial-of-service attacks to be out-of-scope.
Due to this we do not include for example attacks such as SGX-Bomb~\cite{Jang2017}, where the \kw{CPU} can be forced to shutdown due to a hardware bug called RowHammer~\cite{Kim2014a, Seaborn2015}.

The security of SGX enclaves can be undermined if enclaves do not take care to strictly adhere to an agreed-upon secure interface between the trusted and untrusted code bases.
To help facilitate this, a number of \kw{SDK}s are available, to aid the construction of secure enclaves.
In particular they help out with, for example, CPU status flag sanitation, correct stack pointer restoration, range checks of pointers and arrays and prevention of register leakage on exit.
Of course, any vulnerability in the SDK itself will automatically impact all enclaves that make use of it.
In ``A Tale of Two Worlds''~\cite{VanBulck2019a} Van Bulck et al.\ discovered multiple vulnerabilities in all open source SDKs for enclave development that they tested.
This paper do not invalidate the security properties of SGX in and of itself, but it highlights the difficulty in writing secure software in general and enclaves in particular.

Another class of attacks are those that targets specific vulnerabilities of the enclave developer's own implementation.
We mention a few such attacks here, for the readers convenience, but the list is by no means complete.

Checkoway and Shacham proposed the Iago attack~\cite{Checkoway}.
The authors proposed to take advantage of the implicit trust applications place on the kernel, despite it being explicitly stated to exist outside the \kw{TCB}.
Conceptually the Iago attack uses a malicious \kw{OS}-kernel to send false responses to system calls in order to fool the application under attack to perform operations against its own interests.

Lee et al.\ evaluated the old attack technique ``Return-Oriented Programming'' (\kw{ROP}) in~\cite{Lee2017a} and found that it is indeed possible to circumvent the hardware protections provided by the \kw{SGX} design and achieve a total security break of the attacked enclave.

An attack was also introduced by Weichbrodt et al.~\cite{Weichbrodt2016} which exploits synchronization bugs in multi-threaded \kw{SGX} enclaves.
The authors built an attack tool called ``AsyncShock'' which simplifies the reliable exploitation of such bugs in enclave code.

\subsection{Controlled Channel Attacks~\citeNA{Xu2015, VanBulck2017, VanBulck2017a, VanBulck2018b, Gyselinck2018, Wang2017}}\label{sec:controlled_channel}

This section introduces the notion of \emph{controlled channel} attacks, a term that Xu et al.~\cite{Xu2015} coined in 2015.
It is a type of side-channel attack that make use of the near-total control the untrusted \kw{OS}-kernel has over the platform.
This control can be used to construct powerful side channels against the protected enclave who relies on the kernel's services.
There have been several more attacks, based on the same principles.

The strategy used in~\cite{Xu2015,VanBulck2017} was to monitor memory accesses with page-level granularity by monitoring or introducing page-faults.
SGX-Step~\cite{VanBulck2017a} instead configures \kw{APIC} timers, issues interrupts and tracks page-table entries in such a way that it allows for single-stepping enclave code instructions.

SGX-Step is also used as a framework in many other attacks (see~\citeNA{Huo, Murdock2020,Gyselinck2018, VanBulck2018a,VanBulck2018b,VanBulck2019a}).

In~\cite{VanBulck2018b} a timing-based side channel with instruction-level granularity is achieved by timing carefully synchronized interrupts while the enclave is running.
The authors in~\cite{Gyselinck2018} propose a side channel by exploiting the memory segmentation feature only available for code running in the 32-bit legacy mode.

Wang et al.\ proposed a new attack called \emph{sneaky page monitoring}~\cite{Wang2017} which does not require any interrupts of the enclave by periodically accessing and resetting the \emph{accessed} flag in the translation lookaside buffer, \kw{TLB}.





\subsection{Cache-attacks~\citeNA{Moghimi2017a, Brasser2017, Schwarz2017,Gtzfried2017a, Dall2018a, Moghimi2019}}\label{sec:cache_attacks}

There have been many cache-based timing attacks against \kw{SGX} enclaves published in the literature.
Common to all of them are the exploitation of the cache-hierarchy system and the fact that the caching of memory loads from \kw{DRAM} leaves effects in the system state which are measurable from outside the protected application.
What these attacks show is that \kw{SGX} enclaves are vulnerable to the same cache attacks against secret dependent information processing as any software application. In-fact they appear to be even more vulnerable due to the increased capability of the attackers in SGX's attack model.

There exist a number of different general (non \kw{SGX}-specific) techniques for extracting information from side channels, we mention here \emph{Flush+\allowbreak{}Reload}~\cite{Yarom2014a}, \emph{Prime+\allowbreak{}Probe}~\cite{Osvik2006}, \emph{Evict+\allowbreak{}Time}~\cite{Tromer2010a}, \emph{Evict+\allowbreak{}Reload}~\cite{Gruss2015} and \emph{Flush+\allowbreak{}Flush}~\cite{Gruss2016a}.

In the \emph{sneaky page monitoring}\footnote{The basic attack is a controlled channel attack, see~\secref{sec:controlled_channel}.} Wang et al.~\cite{Wang2017} explores several different ways of improving their attack. The authors particularly makes use of the Prime+Probe cache timing technique to increase the granularity of their attack.

The \emph{CacheZoom} attack~\cite{Moghimi2017a}, introduced by Moghimi et al.\ also makes use of the Prime+Probe technique, as does Götzfried et al.~\cite{Gtzfried2017a} in their attack.
Schwarz et al.~\cite{Schwarz2017} construct a malicious enclave from which they mount a Prime+Probe attack against other enclaves.
Brasser et al.\ proposed a same-core attack (using \kw{HT}) against the L1 cache~\cite{Brasser2017}, also using Prime+Probe.
Prime+Probe is again used by Dall et al.~\cite{Dall2018a} to attack Intel's \emph{provisioning enclave} and thereby allows Intel themselves to break \kw{EPID}'s\footnote{EPID or Enhanced Privacy ID is Intel's recommended algorithm used for attestation while preserving privacy of the trusted system } unlinkability property.

The \emph{MemJam}~\cite{Moghimi2019} attack by Moghimi et al.\ uses read-after-write false dependencies due to the \kw{4K} aliasing of the \kw{L1} cache. The methodology itself resembles that of Evict+Time.




\subsection{Branch Prediction Attacks~\citeNA{Lee2017,Evtyushkin2018, Huo}}\label{sec:branch_prediction}

Lee et al.\ introduced the Branch Shadowing attack in~\cite{Lee2017} to reveal fine-grained control flow of a running \kw{SGX} enclave, they showed that this could be used to break the security of several enclave-based constructs.
In~\cite{Evtyushkin2018} Evtyushkin et al.\ proposed a similar attack, dubbed BranchScope, which uses the directional branch predictor instead of the \kw{BTB} (branch target buffer) which is a companion component to the \kw{BTB}.
This shows that the branch predictor unit can be vulnerable even in the face of \kw{BTB} protections.
Bluethunder~\cite{Huo} is another branch predictor attack similar to BranchScope.
The main difference of Bluethunder to BranchScope is that Bluethunder uses a 2-level directional predictor which is a completely different branch predictor unit.
As a result Bluethunder is 52 times faster than BranchScope.


\subsection{Speculative Execution Attacks~\citeNA{Chen2019a,Koruyeh2018,OKeeffe2018}}\label{sec:attacks_spec_exec}

Early 2018 the Spectre~\cite{Kocher2019} and Meltdown~\cite{Lipp2018} attacks made headlines outside the academic world.
In this section, the Spectre attack in particular is of interest.
This attack originally had 2 variants: bounds check bypass and branch target injection.
The second variant targets the \kw{BTB}, branch target buffer, in such a way that when the victim process executes an indirect branch instruction it mispredicts and speculatively executes code that never would have been executed.
Of course, once the \kw{CPU}-pipeline catches up and realizes that it was a misprediction it discards any results.
Central to the Spectre attack however, is the hardware vulnerability that these speculatively executed instructions results in measurable changes to the \kw{CPU} state, or in this case the \kw{CPU} cache.
In short, the Spectre attack allows an attacking process to infer some data values from vulnerable co-hosted processes.

In~\cite{Chen2019a} Chen et al.\ answered the question of whether or not \kw{SGX} is vulnerable to the Spectre or Spectre-like attacks in the affirmative.
The authors presented the SgxPectre attack and used it to extract the secret seal keys and attestation keys from Intel signed quoting enclaves.
In~\cite{Koruyeh2018} Koruyeh et al.\ proposed the SpectreRSB attack which alternatively uses the return stack buffer which is a structure in modern CPUs used to speculatively predict the return address of execution frames (functions).
SgxSpectre in~\cite{OKeeffe2018} (not to be confused with SgxPectre~\cite{Chen2019a}) also demonstrated a successful attack on \kw{SGX} enclaves using a slight modification of the Spectre variant 1 attack.



\subsection{Rogue Data Cache Loads~\citeNA{VanBulck2018a}}\label{sec:rogue_data_cache_loads}

Similarly to the Spectre-type of attacks Meltdown~\cite{Lipp2018} exploits the out-of-order execution of modern CPUs.
Unlike Spectre however, Meltdown does not explicitly make use of the speculative execution feature or the \kw{BTB}, instead it relies on a race condition where the results of unauthorized memory accesses are transiently available for out-of-order executed instructions before the \kw{CPU} issues a fault and rolls-back the results of these instructions.
This implicitly affects the \kw{CPU} cache and allows the memory access to be inferred.

\kw{SGX} however, works slightly differently in that it does not issue any faults for accessing enclave memory, but instead uses \emph{abort page semantics}~\cite{costan2016intel} which allows the access with a dummy -1 result.
The Foreshadow attack~\cite{VanBulck2018a} works around this lack of race condition by instead relying on the fact that the abort page semantics applies only after normal page-table based permission checks succeeds without issuing a page-fault.
Foreshadow therefore revokes all access to the enclave pages that it wishes to read using the \kw{mprotect} system call, after this the principle of the Meltdown approach can again be used.
Due to details of the \kw{SGX} implementation this only works for memory which has been already cached in the first level cache (\kw{L1}), Intel therefore categorized the Foreshadow vulnerability as a ``\kw{L1} Terminal Fault'' (\kw{L1TF}) bug.~\cite{VanBulck2018a}

In Foreshadow-NG~\cite{Weisse2018} the authors generalize the \kw{L1TF} attack into three versions: Foreshadow-SGX (original Foreshadow attack), Foreshadow-OS and Foreshadow-VMM\@.
These attacks were also used in~\cite{VanBulck2019}.
The latter 2 attacks were also discovered by Intel and are not applicable to \kw{SGX}.

\subsection{Microarchitectural Data Sampling (\kw{MDS})~\citeNA{VanSchaik2019, Schwarz2019, VanSchaik2020, Ragab2021}}\label{sec:attacks_mds}

In late 2019 three papers of similar nature were published, namely \emph{Fallout}~\cite{Canella2019}, \emph{RIDL}~\cite{VanSchaik2019} and \emph{Zombie\-Load}~\cite{Schwarz2019}.
Intel has dubbed this new class of attacks \emph{Microarchitectural Data Sampling} attacks, or \kw{MDS}-attacks and they can be used to bypass most of the common security boundaries, such as: JavaScript sandboxes, processes, kernels, VMs and \kw{SGX} enclaves.
Although similar in nature to both Spectre and Meltdown, due to their use of out-of-order and speculative execution features they have another common theme;
these works are based upon leakage of information from a number of implementation specific and undocumented intermediary buffers of the targeted micro architecture.
Closely following these original papers \emph{CacheOut}~\cite{VanSchaik2020} \emph{CrossTalk}~\cite{Ragab2021} and \emph{SXAxe}~\cite{VanSchaik2020b} were published in the first half of 2020.

Fallout, for example, makes use of the \emph{store buffer} to leak information of kernel writes to user space.
Luckily, due to the flushing of the store buffer, \kw{SGX} is safe from this particular attack.
Unluckily, \kw{SGX} still falls victim to the other attacks mentioned in the above paragraph.

RIDL, or \emph{Rogue In-Flight Data Load}, and CacheOut both exploit the \emph{Line Fill Buffers} (\kw{LFB}s) to the effect of bypassing all --- at their respective time of publication --- deployed mitigations for Spectre, Meltdown and Foreshadow.
The \kw{LFB}s are used in the transfer paths between the \kw{L1} data caches (\kw{L1D}) and the \kw{L2} caches.
RIDL was the first work to critically analyse the behaviour of the \kw{LFB}s from a \kw{MDS}-attack perspective.

ZombieLoad also targets the \kw{LFB}s\footnote{at least in part, although the authors speculate on other sources of the leakage as well}, as the name suggests however, it leaks whatever \emph{stale} data currently resides in the buffers.
ZombieLoad is an improvement in that while RIDL only leaks data from loads \emph{not currently} residing in the \kw{L1D} cache, ZombieLoad leaks the results of memory loads, regardless of the requested data's presence in \kw{L1D} or not.

Both RIDL and ZombieLoad suffer from a ``drinking from a firehose''~\cite{VanSchaik2020} problem, in that they are unable to control which data is loaded into the \kw{LFB}s and subsequently leaked.
CacheOut later solved this problem by forcing contention on the \kw{L1D} data it wishes to target, and thereby evicting it (through the \kw{LFB}s) from the cache.

SGAxe~\cite{VanSchaik2020b} is not a new attack, per se, but it utilized the CacheOut attack to extract the sealing key and in turn the machine's attestation key from the Intel provided Quoting Enclave. This key could be used to forge attestation quotes.

CrossTalk~\cite{Ragab2021} extends the \kw{MDS} techniques to show that enclaves can be attacked even across different execution units (i.e.\ cross core). An otherwise good mitigation strategy is to isolate execution of sensitive operation to \kw{CPU} cores not shared with untrusted threads.
CrossTalk achieves this by managing to sample the so called staging buffer which is an undocumented component on modern Intel \kw{CPU}s, shared between all cores.
The paper demonstrates the attack by showing how snooping on the output from the \kw{rdrand} instructions can be used to extract a \kw{ECDSA} private key from a \kw{SGX} enclave running on an separate core.

\subsection{Software-based Fault Injection Attacks~\citeNA{Murdock2020}}\label{sec:soft-fault-injection}

In 2017 the CLKscrew~\cite{Tang2017} by Tang et al.\ attacked ARM TrustZone by adjusting the dynamic frequency scaling, through a privileged and model-specific interface (used by system software for dynamic overclocking).
This attack does not affect Intel \kw{SGX} since it is specific to ARM based systems, but recently Murdock et al.\ published the Plundervolt attack~\cite{Murdock2020} which does affect \kw{SGX}.

The Plundervolt attack abuses privileged interfaces for dynamic voltage scaling on the x86 \kw{CPU} in order to reliably corrupt enclave computations.
The authors write ``Using this interface to very briefly decrease the \kw{CPU} voltage during a computation in a victim \kw{SGX} enclave, we show that a privileged adversary is able to inject faults into protected enclave computations''.
The authors then proceeds to demonstrate how this attack can be used to ``reconstruct full cryptographic keys with negligible computational effort''.~\cite{Murdock2020}

The SGX-Bomb~\cite{Jang2017} and RowHammer~\cite{Kim2014a, Seaborn2015} hardware bugs mentioned earlier would also fit in this category, if denial of service attacks had not been placed out of scope for this article.

\begin{table*}
\centering
\caption{
	Summary of all cited attacks listed here with a number of properties displayed in a table-format.
	Here we use \MatrixYes, \MatrixPartly\ and \MatrixNo\ to mean yes, partly and no, respectively.
}\label{tab:attack_matrix}
\begingroup
\setlength{\tabcolsep}{5pt} 
\begin{tabular}{@{}llllllllllllllll@{}}
	\multicolumn{1}{c}{}                             & \multicolumn{1}{c}{}           & \multicolumn{1}{c}{}                   &                       &  & \multicolumn{6}{c}{Impact} &                         & \multicolumn{4}{c}{Mitigations}                                                                                                                                                                                                                           \\ \cmidrule(lr){6-11} \cmidrule(l){13-16}
	\multicolumn{1}{c}{Attacks (abbreviated titles)} & \multicolumn{1}{c}{\rot{Type}} & \multicolumn{1}{c}{\rot{SGX Specific}} & \rot{Targeted attack} &  & \rot{Page access pattern}  & \rot{Instruction trace} & \rot{Instruction latency}       & \rot{Memory access pattern} & \rot{Memory Contents} & \rot{Fault Injection} &  & \rot{Microcode patch}         & \rot{System design}      & \rot{Compiler/SDK}                             & \rot{Application design}   \\ \midrule
	Controlled-Channel~\cite{Xu2015}                 & \AttackControlledChannel{}     & \MatrixYes{}                           & \MatrixYes{}          &  & \MatrixYes{}               & \MatrixNo{}             & \MatrixNo{}                     & \MatrixNo{}                 & \MatrixNo{}           & \MatrixNo{}           &  & \MatrixNo{}                   & \MatrixNo{}              & \MatrixYesC{Strackx2017}                       & \MatrixYesC{Shinde2015}    \\
	Stealthy Page Table~\cite{VanBulck2017}          & \AttackControlledChannel{}     & \MatrixYes{}                           & \MatrixYes{}          &  & \MatrixYes{}               & \MatrixNo{}             & \MatrixNo{}                     & \MatrixNo{}                 & \MatrixNo{}           & \MatrixNo{}           &  & \MatrixNo{}                   & \MatrixNo{}              & \MatrixYesC{Strackx2017}                       & \MatrixNo{}                \\
	SGX-Step~\cite{VanBulck2017a}                    & \AttackControlledChannel{}     & \MatrixYes{}                           & \MatrixNo{}           &  & \MatrixYes{}               & \MatrixYes{}            & \MatrixYes{}                    & \MatrixNo{}                 & \MatrixNo{}           & \MatrixNo{}           &  & \MatrixNo{}                   & \MatrixNo{}              & \MatrixYesCC{VanBulck2017a}{Hosseinzadeh2018}  & \MatrixNo{}                \\
	Nemesis~\cite{VanBulck2018b}                     & \AttackControlledChannel{}     & \MatrixYes{}                           & \MatrixYes{}          &  & \MatrixNo{}                & \MatrixNo{}             & \MatrixYes{}                    & \MatrixNo{}                 & \MatrixNo{}           & \MatrixNo{}           &  & \MatrixNo{}                   & \MatrixNo{}              & \MatrixYesC{Shih2017}                          & \MatrixNo{}                \\
	Off Limits~\cite{Gyselinck2018}                  & \AttackControlledChannel{}     & \MatrixYes{}                           & \MatrixYes{}          &  & \MatrixYes{}               & \MatrixPartly{}         & \MatrixNo{}                     & \MatrixPartly{}             & \MatrixNo{}           & \MatrixNo{}           &  & \MatrixYesC{Gyselinck2018}    & \MatrixNo{}              & \MatrixNo{}                                    & \MatrixYesC{Gyselinck2018} \\
	Leaky Cauldron~\cite{Wang2017}                   & \AttackCache{}                 & \MatrixYes{}                           & \MatrixYes{}          &  & \MatrixYes{}               & \MatrixNo{}             & \MatrixNo{}                     & \MatrixYes{}                & \MatrixNo{}           & \MatrixNo{}           &  & \MatrixNo{}                   & \MatrixNo{}              & \MatrixYesCC{Shih2017}{Chen2017}               & \MatrixNo{}                \\
	CacheZoom~\cite{Moghimi2017a}                    & \AttackCache{}                 & \MatrixYes{}                           & \MatrixYes{}          &  & \MatrixNo{}                & \MatrixNo{}             & \MatrixNo{}                     & \MatrixYes{}                & \MatrixNo{}           & \MatrixNo{}           &  & \MatrixNo{}                   & \MatrixNo{}              & \MatrixYesCC{Shih2017}{Chen2017}               & \MatrixYesC{Gruss2017a}    \\
	Cache Attacks on \kw{SGX}~\cite{Gtzfried2017a}   & \AttackCache{}                 & \MatrixYes{}                           & \MatrixYes{}          &  & \MatrixNo{}                & \MatrixNo{}             & \MatrixNo{}                     & \MatrixYes{}                & \MatrixNo{}           & \MatrixNo{}           &  & \MatrixNo{}                   & \MatrixNo{}              & \MatrixNo{}                                    & \MatrixYesC{Gruss2017a}    \\
	Malware Guard Extensions~\cite{Schwarz2017}      & \AttackCache{}                 & \MatrixYes{}                           & \MatrixYes{}          &  & \MatrixNo{}                & \MatrixNo{}             & \MatrixNo{}                     & \MatrixYes{}                & \MatrixNo{}           & \MatrixNo{}           &  & \MatrixNo{}                   & \MatrixYesC{Schwarz2017} & \MatrixNo{}                                    & \MatrixNo{}                \\
	Software Grand Exposure~\cite{Brasser2017}       & \AttackCache{}                 & \MatrixYes{}                           & \MatrixYes{}          &  & \MatrixNo{}                & \MatrixNo{}             & \MatrixNo{}                     & \MatrixYes{}                & \MatrixNo{}           & \MatrixNo{}           &  & \MatrixNo{}                   & \MatrixNo{}              & \MatrixNo{}                                    & \MatrixYesC{Brasser2017}   \\
	CacheQuote~\cite{Dall2018a}                      & \AttackCache{}                 & \MatrixYes{}                           & \MatrixYes{}          &  & \MatrixNo{}                & \MatrixNo{}             & \MatrixNo{}                     & \MatrixYes{}                & \MatrixNo{}           & \MatrixNo{}           &  & \MatrixNo{}                   & \MatrixYesC{Dall2018a}   & \MatrixNo{}                                    & \MatrixNo{}                \\
	MemJam~\cite{Moghimi2019}                        & \AttackCache{}                 & \MatrixNo{}                            & \MatrixYes{}          &  & \MatrixNo{}                & \MatrixNo{}             & \MatrixNo{}                     & \MatrixYes{}                & \MatrixNo{}           & \MatrixNo{}           &  & \MatrixNo{}                   & \MatrixNo{}              & \MatrixYesC{Chen2018}                          & \MatrixNo{}                \\
	Branch Shadowing~\cite{Lee2017}                  & \AttackBranchPredict{}         & \MatrixYes{}                           & \MatrixYes{}          &  & \MatrixNo{}                & \MatrixYes{}            & \MatrixNo{}                     & \MatrixNo{}                 & \MatrixNo{}           & \MatrixNo{}           &  & \MatrixNo{}                   & \MatrixNo{}              & \MatrixYesCC{Lee2017}{Hosseinzadeh2018}        & \MatrixNo{}                \\
	BranchScope~\cite{Evtyushkin2018}                & \AttackBranchPredict{}         & \MatrixNo{}                            & \MatrixYes{}          &  & \MatrixNo{}                & \MatrixYes{}            & \MatrixNo{}                     & \MatrixNo{}                 & \MatrixNo{}           & \MatrixNo{}           &  & \MatrixNo{}                   & \MatrixNo{}              & \MatrixYesCC{Evtyushkin2018}{Hosseinzadeh2018} & \MatrixNo{}                \\
	Bluethunder~\cite{Huo}                           & \AttackBranchPredict{}         & \MatrixYes{}                           & \MatrixYes{}          &  & \MatrixNo{}                & \MatrixYes{}            & \MatrixNo{}                     & \MatrixNo{}                 & \MatrixNo{}           & \MatrixNo{}           &  & \MatrixNo{}                   & \MatrixNo{}              & \MatrixYesC{Huo}                               & \MatrixNo{}                \\
	SgxPectre~\cite{Chen2019a}                       & \AttackSpeculative{}           & \MatrixYes{}                           & \MatrixYes{}          &  & \MatrixNo{}                & \MatrixNo{}             & \MatrixNo{}                     & \MatrixNo{}                 & \MatrixYes{}          & \MatrixNo{}           &  & \MatrixYesC{Canella2018}      & \MatrixNo{}              & \MatrixNo{}                                    & \MatrixNo{}                \\
	SpectreRSB~\cite{Koruyeh2018}                    & \AttackSpeculative{}           & \MatrixNo{}                            & \MatrixYes{}          &  & \MatrixNo{}                & \MatrixNo{}             & \MatrixNo{}                     & \MatrixNo{}                 & \MatrixYes{}          & \MatrixNo{}           &  & \MatrixNo{}                   & \MatrixNo{}              & \MatrixNo{}                                    & \MatrixYesC{Koruyeh2018}   \\
	Spectre v1~\cite{OKeeffe2018}                    & \AttackSpeculative{}           & \MatrixNo{}                            & \MatrixYes{}          &  & \MatrixNo{}                & \MatrixNo{}             & \MatrixNo{}                     & \MatrixNo{}                 & \MatrixYes{}          & \MatrixNo{}           &  & \MatrixYesC{Canella2018}      & \MatrixNo{}              & \MatrixNo{}                                    & \MatrixNo{}                \\
	Foreshadow-SGX~\cite{VanBulck2018a}              & \AttackRogueDataCacheLoad{}    & \MatrixYes{}                           & \MatrixNo{}           &  & \MatrixNo{}                & \MatrixNo{}             & \MatrixNo{}                     & \MatrixNo{}                 & \MatrixYes{}          & \MatrixNo{}           &  & \MatrixYesC{Weisse2018}       & \MatrixNo{}              & \MatrixNo{}                                    & \MatrixNo{}                \\
	RIDL~\cite{VanSchaik2019}                        & \AttackMDS{}                   & \MatrixNo{}                            & \MatrixYes{}          &  & \MatrixNo{}                & \MatrixNo{}             & \MatrixNo{}                     & \MatrixNo{}                 & \MatrixYes{}          & \MatrixNo{}           &  & \MatrixYesC{VanSchaik2019}    & \MatrixNo{}              & \MatrixNo{}                                    & \MatrixNo{}                \\
	ZombieLoad~\cite{Schwarz2019}                    & \AttackMDS{}                   & \MatrixNo{}                            & \MatrixYes{}          &  & \MatrixNo{}                & \MatrixNo{}             & \MatrixNo{}                     & \MatrixNo{}                 & \MatrixYes{}          & \MatrixNo{}           &  & \MatrixYesC{Schwarz2019}      & \MatrixNo{}              & \MatrixNo{}                                    & \MatrixNo{}                \\
	CacheOut~\cite{VanSchaik2020}                    & \AttackMDS{}                   & \MatrixYes{}                           & \MatrixYes{}          &  & \MatrixNo{}                & \MatrixNo{}             & \MatrixNo{}                     & \MatrixNo{}                 & \MatrixYes{}          & \MatrixNo{}           &  & \MatrixPartlyC{VanSchaik2020} & \MatrixNo{}              & \MatrixNo{}                                    & \MatrixNo{}                \\
	CrossTalk~\cite{Ragab2021}                       & \AttackMDS{}                   & \MatrixYes{}                           & \MatrixNo{}           &  & \MatrixNo{}                & \MatrixNo{}             & \MatrixNo{}                     & \MatrixNo{}                 & \MatrixYes{}          & \MatrixNo{}           &  & \MatrixYesC{Ragab2021}        & \MatrixNo{}              & \MatrixNo{}                                    & \MatrixNo{}                \\
	Plundervolt~\cite{Murdock2020}                   & \AttackSoftFaultInject{}       & \MatrixYes{}                           & \MatrixPartly{}       &  & \MatrixNo{}                & \MatrixNo{}             & \MatrixNo{}                     & \MatrixNo{}                 & \MatrixNo{}           & \MatrixYes{}          &  & \MatrixYesC{Murdock2020}      & \MatrixNo{}              & \MatrixNo{}                                    & \MatrixNo{}                \\ \bottomrule
\end{tabular}
\endgroup
\end{table*}

\section{Defensive Strategies and Mitigations}\label{sec:mitigate}

In this section, we present the most relevant published mitigation techniques for the presented attacks and place each into the categories: \DefenceTypeMicroCode, \DefenceTypeSystem, \DefenceTypeCompiler\ and \DefenceTypeApplication.

\paragraph{Defences that are out of scope} This survey does not account for purely theoretical defenses and mitigations which relies on changes to hardware and \kw{ISA}. Some examples are Sanctum~\cite{Costan2016} and Autarky~\cite{Orenbach2020} which both proposes changes to the hardware and \kw{ISA}.

\subsection{\DefenceTypeMicroCode}\label{sec:def-microcode}

A CPU is not fully realized in hardware but rather most of the more complex instructions are implemented by a form of low-level software called microcode.
The microcode can only be changed by the manufacturer of the CPU, in this case Intel.
A microcode patch can thus make direct changes in how the CPU performs its duties, and these patches are usually the most effective way to mitigate any vulnerability.


\subsection{\DefenceTypeSystem}\label{sec:def-system}

Some attacks cannot easily be thwarted by microcode patches but may instead be fixed by redesigning or removing implementation issues of the supporting systems.
For example, Intel provides a number of special enclaves to support higher level services such as the \emph{Launcher Enclave} (\kw{LE}), \emph{Provisioning Enclave} (\kw{PE}) and \emph{Quoting Enclave} (\kw{QE}).

Some microcode patches include new information of the running system, which the attestation services and supporting enclaves must act upon, or otherwise include in the attestation reports~\cite{Canella2018}.

\subsection{\DefenceTypeCompiler}\label{sec:def-compiler}

In some cases Intel does not appear to provide any solutions or mitigations for attacks, but leaves the responsibility up to the enclave authors themselves.
In this case the best one can hope for is for some kind of general approach implemented in either the compiler or in the enclave \kw{SDK}.
By going in this direction its use is, in theory, transparent to the enclave developer. That-is, if they elect to opt-in to the techniques and their various performance impacts and drawbacks.
In this section, we will explore some known solutions that strive to fit in this category.

Lee et al.~\cite{Lee2017} proposed \emph{ZigZagger} as a defence against their own branch shadowing attack.
It works by transforming conditional branches into unconditional jumps to intermediate code sections (called trampolines) that in-turn bounces to the target code.
Hosseinzadeh et al.~\cite{Hosseinzadeh2018} later improves upon this idea by doing control flow randomization at run-time.
Both of these schemes were implemented as compiler extensions on top of \kw{LLVM}.
Meanwhile Chen et al.\ presents a solution also implemented in \kw{LLVM} that closes \kw{HT} (Hyper-Threading) based side-channels~\cite{Chen2018} by blocking access to sibling cores via the creation of a shadow thread\footnote{
	Previously it would not help to disable Hyper-Threading since its status was not included in the attestation reports.
	But with the introduction of the Foreshadow mitigations, disabling \kw{HT} in \kw{BIOS} would now appear to be a more secure alternative.
}.

Shih et al.\ presented a modified \kw{LLVM} compiler dubbed T-SGX~\cite{Shih2017} which terminates the execution of sensitive operations using Intel’s Transactional Synchronization Extensions, \kw{TSX}, if a certain number of interrupts occurs.
T-SGX is claimed to be effective against all known controlled channel attacks.
``Déjà Vu'' is an alternative solution proposed by Chen et al.\ in~\cite{Chen2017} which implements a clock-thread protected with \kw{TSX} by which it can be detected if the run-time of the program differs significantly from what is expected.
If it does, it is assumed that the protected code has been interrupted and Déjà Vu will therefore abort the execution.

Strackx et al.\ presented the ``Heisenberg Defence''~\cite{Strackx2017} as an alternative.
It also utilizes \kw{TSX} transactions, but by adding preloading and verifications it can proactively protect sensitive code against page-table based controlled channel attacks.
For systems without \kw{TSX} support, SGX-LAPD was proposed by Fu et al.\ in~\cite{Fu2017} using a detection based approached similar to T-SGX but instead implemented using large pages.

To defeat enclave specific attacks such as, for example, \kw{ROP} attacks (which remain out of scope for this paper) Seo et al.\ \cite{Seo2017} found a way to activate \kw{ASLR} inside SGX enclaves, to make exploitation more difficult.
It is implemented on top of the \kw{LLVM} compiler.


\subsection{\DefenceTypeApplication}\label{sec:def-application}

If all else fails, the enclave's author must take care to design their enclaves in a secure and side-channel protected manner.
In this section we discuss some of the published tools and techniques that have been proposed to help with this.

Shinde et al.\ suggest a compiler \emph{assisted} solution to remove data dependent memory accesses~\cite{Shinde2015} in order to mitigate page-fault based side channels by aiming to keep secret data and code within the same page.
Cloak is a software library developed by Gruss et al.\ in~\cite{Gruss2017a} that allows secret data and secret handling code to be wrapped in a \kw{TSX} transaction.
It seems to be quite effective at preventing cache-based side-channels.

An alternative solution is to introduce random noise in the applications algorithms by adding accesses to dummy data.
Chandra et al.\ explores this possibility in~\cite{Chandra2017}.
The same effect can be more thoroughly obtained by the use of \emph{Oblivious RAM} (\kw{ORAM}) constructions with \emph{data oblivious execution} to hide memory accesses in such a way that none-of the above mentioned side-channel attacks would be effective.
ZeroTrace~\cite{Sasy2018a} introduced by Sasy et al.\ is one such scheme for use in \kw{SGX} enclaves.


\section{Discussion on the current status of mitigations}\label{sec:status}


In table~\ref{tab:attack_matrix}, we have compiled a matrix that summarizes a number of properties for each attack, including the current state of mitigations, to the best of our knowledge.

As discussed earlier, the mitigation techniques are divided into four categories: \DefenceTypeMicroCode, \DefenceTypeSystem, \DefenceTypeCompiler\ and \DefenceTypeApplication.
As it can be seen in Table~\ref{tab:attack_matrix}, almost all categorized attacks are mitigated or partially mitigated by these techniques.

Usually, the mitigation techniques for attacks in the same category are similar to each other.
For instance, Controlled Channel Attacks~\cite{Xu2015, VanBulck2017, VanBulck2017a, VanBulck2018b, Gyselinck2018, Wang2017} and Branch Prediction Attacks~\cite{Lee2017,Evtyushkin2018, Huo} can be mitigated using \DefenceTypeCompiler\ techniques or modifying the \DefenceTypeApplication.

The Cache-attacks~\citeNA{Moghimi2017a, Brasser2017, Schwarz2017,Gtzfried2017a, Dall2018a, Moghimi2019} are explicitly outside the scope of the Intel threat-model. This, in combination with the fact that these attack are very specific for each targeted enclave implementation is the reason why one would be forced to make enclave specific mitigations (i.e.\ modifying the application design).
In the case of the cache attack in~\cite{Dall2018a}, since it directly attacks part of the attestation service, the only viable option is for Intel to update the affected enclaves. At this time we have found no indication whether or not this has been done.

Most of the Speculative Execution Attacks, Rogue Data Cache Loads, \kw{MDS} Attacks, and Software-based Fault Injection Attacks are mitigated using \DefenceTypeMicroCode.
This is the case for the 32-bit memory segmentation attack ``Off-Limits'' in~\cite{Gyselinck2018}, SgxPectre~\cite{Chen2019a}, Foreshadow~\cite{VanBulck2018a}, Fores\-ha\-dow-NG~\cite{Weisse2018} and PlunderVolt~\cite{Murdock2020}. In a few instances the mitigation is in combination with updates to Intel provided attestation services~\cite{Canella2018}.

While many of these mitigations appear to be quite effective, in practice some of them impose additional requirements on the system.
For example, part of the mitigation strategy against Foreshadow would be to disable Hyper-Threading (\kw{HT}) (logical cores share the same \kw{L1} cache) and for this reason the status of \kw{HT} is included during attestation and sealing operations.
These mitigations have been supplied through microcode-updates.

Koruyeh et al.\ mentions in SpectreRSB paper~\cite{Koruyeh2018} that the microcode patch ``RSB-refilling'' is not specifically applied for \kw{SGX} enclaves.
Based on this we draw the conclusion that RSB-refilling ought to be implemented either the SDKs or compilers, if they wish to ensure protection against the SpectreRSB\@.
We have been unable find any information on whether or not this or some other equivalent mitigation is implemented in the available \kw{SDK}s.
Alternatively, the enclave authors might wish to add this protection manually (this is shown in the table under the ``Application Design'' mitigation strategy).

For most \kw{MDS} type of attacks,~\cite{VanSchaik2019,Schwarz2019} Intel microcode-patches had been already released, but for a more recent attack of this type called CacheOut~\cite{VanSchaik2020} the update patch had not yet been released, at the time of this writing.
Intel has promised~\cite{VanSchaik2020} that the patch will be released in the near future. Until then, we mark this attack as partially mitigated in Table~\ref{tab:attack_matrix}.

The \kw{BIOS} patch and microcode updates which mitigates PlunderVolt works by disabling much of the dynamic voltage scaling interface as well as recording and verifying the status of these interfaces into its sealing and attestation operations.

\section{Conclusions}\label{sec:conclusions}

We found more than 20 attacks in the literature using side channels and active attackers, see Table~\ref{tab:attack_matrix}.
For all of the attacks there are mitigation strategies, however some are more feasible in practice than others. Especially those mitigations which are based on a correct application design might, in practice, be difficult to implement completely.

The number of side channel attacks targeting \kw{SGX} found recently hint that there might be more attacks not yet discovered.
This should also be considered when evaluating if and how to employ Intel \kw{SGX} as a protection mechanism.

\printbibliography{}
\end{document}